\documentclass[prd,onecolumn,superscriptaddress,showpacs,floatfix,%
preprintnumbers, nofootinbib,hyperref]{revtex4}              
\usepackage{graphicx}
\usepackage{rotating}
\usepackage{epsfig}
\usepackage[usenames,dvipsnames]{xcolor}

\usepackage{color}
\def\neff{N_{\rm eff}}
\def\he4{$^4$He}
\def\h2{$^2$H}

\def\P{{\mathcal P}}

\begin{document}

\preprint{DF/2013-3, LAPTH-004/13}
\title{Multi-momentum and multi-flavour  active-sterile neutrino oscillations in the early universe: role of neutrino
 asymmetries and effects on nucleosynthesis}

\author{Ninetta Saviano} 
\affiliation{II Institut f\"ur Theoretische Physik, Universit\"at Hamburg, Luruper Chaussee 149, 22761 Hamburg, Germany} 
\affiliation{Dipartimento di Scienze Fisiche, Universit{\`a} di Napoli Federico II,
Complesso Universitario di Monte S. Angelo, I-80126 Napoli, Italy}
\affiliation{Istituto Nazionale di Fisica Nucleare - Sezione di Napoli, Complesso Universitario di Monte S. Angelo, I-80126 Napoli, Italy}
\author{Alessandro Mirizzi} 
\affiliation{II Institut f\"ur Theoretische Physik, Universit\"at Hamburg, Luruper Chaussee 149, 22761 Hamburg, Germany}
\author{Ofelia Pisanti}
 \affiliation{Dipartimento di Scienze Fisiche, Universit{\`a} di Napoli Federico II, Complesso Universitario di Monte S. Angelo, I-80126 Napoli, Italy}
\affiliation{Istituto Nazionale di Fisica Nucleare - Sezione di Napoli, Complesso Universitario di Monte S. Angelo, I-80126 Napoli, Italy}
 \author{Pasquale Dario Serpico} 
\affiliation{LAPTh, Univ. de Savoie, CNRS, B.P.110, Annecy-le-Vieux F-74941, France}
\author{Gianpiero Mangano}
\affiliation{Istituto Nazionale di Fisica Nucleare - Sezione di Napoli, Complesso Universitario di Monte S. Angelo, I-80126 Napoli, Italy}
 \author{Gennaro Miele}
 \affiliation{Dipartimento di Scienze Fisiche, Universit{\`a} di Napoli Federico II, Complesso Universitario di Monte S. Angelo, I-80126 Napoli, Italy}
\affiliation{Istituto Nazionale di Fisica Nucleare - Sezione di Napoli, Complesso Universitario di Monte S. Angelo, I-80126 Napoli, Italy}

\date{\today}

\begin{abstract}
We perform a  study of the flavour evolution in the early universe of a 
multi-flavour  active-sterile neutrino system with parameters inspired by 
the short-baseline neutrino anomalies. In a neutrino-symmetric bath  a ``thermal'' population of the sterile state would quickly grow, but in the presence of primordial neutrino asymmetries a self-suppression as well as a resonant sterile neutrino production can take place, depending on
temperature and chosen parameters. In order to characterize these effects,  we go beyond the usual average momentum and single mixing approximations and consider a multi-momentum and multi-flavour treatment of the kinetic equations. We find that the enhancement obtained in this case with respect to the average momentum approximation is significant, up to $\sim 20\%$ of a degree of freedom. Such detailed and computationally demanding treatment further raises the asymmetry values required to significantly suppress the sterile neutrino production, up to  $|L_{\nu}| \gtrsim {\cal O}(10^{-2})$. For such asymmetries, however, the active-sterile flavour conversions happen so late that significant distortions are produced in the electron (anti)neutrino spectra. The larger  $|L_{\nu}|$, the more the impact of these distortions takes over as dominant cosmological effect, notably increasing the $^4$He abundance in primordial nucleosynthesis (BBN). The standard expression of  the primordial yields in terms of the effective number of neutrinos and asymmetries is also greatly altered.
We numerically estimate the magnitude of such effects for a few representative cases and comment on  the implications for current cosmological measurements.

\end{abstract}


\maketitle

\section{Introduction}

In recent years different short-baseline neutrino oscillation experiments have found anomalous results
that may be interpreted by enlarging the standard description of neutrino oscillations in terms
of three active species. In particular, 
the    ${\overline \nu}_\mu \to {\overline\nu}_e$ oscillations in
LSND~\cite{Aguilar:2001ty} and MiniBoone~\cite{AguilarArevalo:2010wv} experiments (recently constrained by the 
ICARUS experiment~\cite{Antonello:2012pq}), the ${\overline\nu}_e$ and $\nu_e$ disappearance
revealed by the Reactor Anomaly~\cite{Mention:2011rk}, and the Gallium Anomaly~\cite{Acero:2007su} can be described
in terms of  light [$m\sim\mathcal O(1)\,$eV] sterile neutrinos which mix with the active ones 
(see~\cite{Abazajian:2012ys,Palazzo:2013me} for  recent reviews). 
In this context, scenarios with one (dubbed ``3+1'') or two (``3+2'')  sterile neutrinos~\cite{Akhmedov:2010vy,Kopp:2011qd,Giunti:2011gz,Giunti:2011hn,Donini:2012tt,Giunti:2012tn,Kopp:2013vaa} have been proposed to fit the different data. 

Cosmological measurements represent a powerful tool to  probe the number of neutrinos and their mass 
at eV scale (see, e.g.,~\cite{Lesgourgues:2006nd,Wong:2011ip,RiemerSorensen:2013ih}). 
The non-electromagnetic radiation content in the universe is usually expressed in terms
of the effective number of excited neutrino species, $N_{\rm eff}$.  This can be constrained by 
 Cosmic Microwave Background (CMB)~\cite{Komatsu:2010fb,Hou:2011ec,Das:2010ga}, 
Large Scale Structure (LSS)~\cite{Hou:2011ec,Das:2010ga}, and Big Bang Nucleosynthesis (BBN) data~\cite{Mangano:2011ar,Hamann:2011ge,Pettini:2012ph}. 
 Current measurements, especially CMB ones, slightly favor the existence of some extra radiation, though the results of WMAP-9 data release~\cite{Hinshaw:2012fq} and Atacama Cosmology Telescope \cite{Sievers:2013wk}, when combined with other cosmological constraints, are compatible with the Standard Model expectation value, $N_{\rm eff}=3.046$~\cite{Mangano:2005cc}. 
In particular, WMAP-9 finds  $N_{\rm eff} = 3.84 \pm 0.40$ when the full data are analyzed~\cite{Hinshaw:2012fq},
 while Atacama gives  $N_{\rm eff} = 2.79 \pm 0.56$.
In this context, a recent Bayesian analysis of the current cosmological datasets does not show  a
strong preference for a value of $N_{\rm eff}$ larger than the  standard one~\cite{verde}.
Very recently, a combination of cosmological data notably including the ones released by Planck has yielded
$N_{\rm eff} = 3.30\pm 0.27$~\cite{Ade:2013lta}.

Similarly, up to about one fully thermalized  sterile neutrino is still marginally allowed by BBN~\cite{Mangano:2011ar,Pettini:2012ph},
while a value corresponding to two sterile states appears largely excluded~\cite{Jacques:2013xr}, but no significant preference for a larger-than-standard value of $\neff$ either. Notice, however, that even a single extra thermalized sterile neutrino with  mass 
$m\sim\mathcal O(1)\,$eV
appears to be inconsistent with mass bounds from CMB and LSS data~\cite{Hamann:2010bk,GonzalezGarcia:2010un,Hamann:2011ge,RiemerSorensen:2012ve,Archidiacono:2012ri,Joudaki:2012uk}.
  
In order to reconcile the eV sterile neutrino interpretation of the short-baseline anomalies with the cosmological
observations, the most straightforward possibility would be  to suppress  the sterile neutrino thermalization in the early universe, correspondingly reducing 
the expected excess in  extra-radiation. 
 In this sense, it was proposed at first in 
in~\cite{Foot:1995bm} (see also~\cite{Chu:2006ua}) to consider a  primordial  asymmetry between neutrinos and antineutrinos~\footnote{Alternative escape
routes have been recently proposed, see e.g.~\cite{ho:2012}.} 
\begin{equation}
L_{\nu} = \frac{n_{\nu}-n_{\bar\nu}}{n_{\gamma}} \,\ . 
\label{eq:leptonasy}
\end{equation}
In principle, one would expect the neutrino asymmetry to be of the same order of magnitude of the baryonic one, 
$\eta = (n_B-n_{\bar B})/n_{\gamma} \simeq 6 \times 10^{-10}$. This of course holds for charged leptons, due to the stringent requirement of charge neutrality of the universe, but not necessarily for neutrinos. In fact,  the constrains on $L_{\nu}$ are quite loose,
allowing also  $|L_{\nu}|\simeq 10^{-2}-10^{-1}$~\cite{Dolgov:2002ab,Wong:2002fa,Abazajian:2002qx,Serpico:2005bc,Pastor:2008ti,Shimon:2010ug,Mangano:2010ei,DiValentino:2011sv,Mangano:2011ip,Castorina:2012md}. 
A primordial neutrino asymmetry would  add an additional ``matter term potential'' in the active-sterile neutrino equations of motion. If sufficiently
large, one expects this term to \emph{block} the active-sterile flavour conversions via the in-medium suppression of the mixing angle.
Nevertheless, this term can also generate  Mikheev-Smirnov-Wolfenstein~\cite{Matt} (MSW)-like resonant flavour conversions  among active and sterile
neutrinos, \emph{enhancing} their production. In order to assess which of the two effects dominates the flavour evolution  it is mandatory to perform a study 
of the kinetic equations for active-sterile neutrino oscillations. 

In a recent paper~\cite{Mirizzi:2012we} some of us performed a first study of active-sterile flavour conversions in the presence
of neutrino asymmetries, considering (3+1) and (2+1) scenarios inspired by the recent fits of all the short-baseline 
neutrino anomalies. In order to simplify the numerical complexity of the problem, we adopted equations of motion integrated over momenta, often referred to as  ``average (or single) momentum approximation''. Loosely speaking, this can be thought of as an approximation in which all neutrinos  share the common thermal average comoving
 momentum $\langle p \rangle / T=3.15$. Under this assumption,
  we found that in the case of equal asymmetries among the active species, a value of $|L_{\nu}| \simeq 10^{-3}$ is required
to start suppressing the resonant sterile production.
An even  larger results,  $|L_{\nu}| \simeq 10^{-2}$, is 
necessary to lower the sterile neutrino abundance in case of opposite initial asymmetries. 

However,  $N_{\rm eff}$
is not the only parameter that can affect the cosmological observables. Indeed, for such  large values of  $L_{\nu}$
 resonant active-sterile neutrino conversions  occur near or after the decoupling temperature for the active neutrinos,
making ineffective the repopulation of active species through collisions. 
The lack of repopulation of electron neutrinos would in general produce distorted distributions which can move up the $n/p$ freeze-out and hence increase 
the ${}^{4} {\rm He}$ yield, the main product of BBN. In order to characterize the possible distortions in the active neutrino spectra,
it is necessary to go beyond the average momentum approximation and consider a detailed treatment 
of the full momentum-dependent kinetic equations, due to the momentum-dependence of the resonant conversions between active and sterile neutrinos.

The  purpose of the present work is to perform for the first time  a full multi-flavour and multi-momentum treatment of the active-sterile neutrino oscillations for 
the  (2+1) scenario considered in~\cite{Mirizzi:2012we} in the presence of primordial neutrino asymmetries.
It was shown in~\cite{Mirizzi:2012we}
that this model captures the main features of the complete (3+1) scenario.
In particular, we follow the evolution of the system in the presence of $|L_{\nu}| \gtrsim 10^{-3}$  
where the distortions of the active neutrino spectra start to become sizable. We  remark that for smaller values
of the active neutrino asymmetries the thermalization of the sterile neutrinos is complete, producing 
a tension with cosmological data.
 In the studied cases, we observed an enhancement
in the sterile neutrino production of up to $\Delta \neff\simeq 0.2$  with respect to what observed in the average momentum study~\cite{Mirizzi:2012we}.
This implies that one needs to consider  even larger asymmetries (at least of the order of $|L_\nu| \sim 10^{-2})$ in order
to significantly suppress the production of sterile neutrinos. 
On the other hand, for these values of the asymmetries we find relevant distortions in the electron (anti)neutrino spectra
 notably modifying  the BBN yields (see, e.g.,~\cite{Kirilova:1996fy,Kirilova:1998jx}). 
Computing reliably  these distortions and $\neff$ as functions of the asymmetry parameters  is a very challenging
task, involving time consuming numerical calculations for the flavour evolution. 
  Nonetheless, the few representative values of the asymmetries we investigated already allow to draw non-trivial implications for cosmological observables.

This article is structured as follows: in Section~II we briefly summarize our formalism, highlighting the modifications with respect to~\cite{Mirizzi:2012we}. We present our results concerning the sterile
neutrino production in the multi-momentum scenario and the differences with respect to the average momentum case. Then, in Section~\ref{BBN} we  show the  distortion of the electron
neutrino spectra for different values of asymmetries and we discuss the impact on observables, namely $N_{\rm eff}$ and $^4$He and $^2$H abundances.
Finally, in Section~\ref{concl} we conclude.

\section{Multi-momentum flavour evolution}

\subsection{Equations of motion}\label{formalism}

In this Section we summarize  the Equations of Motion (EoMs) for the  (2+1) active-sterile neutrino system in the early universe, using 
 the same 
notation of~\cite{Mirizzi:2012we}, to which we address the reader  for details. 
In particular, we describe the time evolution of the neutrino ensemble in terms of
 the following dimensionless variables (replacing time, momentum and photon temperature, respectively)
\begin{equation}
x \equiv m\,a  \qquad  y \equiv p\,a  \qquad z \equiv T_\gamma\,
a~, \label{comoving}
\end{equation}
where $m$ is an arbitrary mass scale which we  put equal to  the electron mass $m_e$. Note that the function $a$ is normalized, without loss of generality, so that $a(t)\to 1/T$ at large temperatures, $T$ being the common temperature of the particles in equilibrium far from any entropy-release process. With this choice, $a^{-1}$ can be identified with the initial temperature of thermal, active neutrinos. 

In order to take into account the interplay between oscillations and collisions of neutrinos, it is necessary to 
describe  the neutrino (antineutrino) system in terms of  $3\times 3$ density matrices $\varrho$  ($\bar\varrho$)\footnote{Here $\nu_{\mu}$ refers generically to a non-electron active  flavour state.}
\begin{equation}
\varrho(x,y) =
\left(\begin{array}{ccc} 
\varrho_{ee} &  \varrho_{e \mu} & \varrho_{es} \\
\varrho_{\mu e}  & \varrho_{\mu \mu} &  \varrho_{\mu s} \\
\varrho_{s e} &\varrho_{s \mu} &\varrho_{ss}
\end{array}\right)~. 
\label{eq:rho}
\end{equation}
In this formalism we write the EoMs    for $\varrho$  and $\bar\varrho$ as~\cite{McKellar:1992ja,Sigl:1992fn,Dolgov:2002ab}
\begin{eqnarray}
i  \frac{d\varrho}{dx} &=& + \frac{x^2 }{2 m^2\,y\,\overline{H}} \left[{\sf M}^2, \varrho \right]
+  \frac{\sqrt{2} G_F \, m^2}{x^2 \, \overline{H}}\left[\left(-\frac{8\, y\, m^2}{3\, x^2 \, m_W^2} {\sf E_\ell}-\frac{8\, y\, m^2 }{3\, x^2 \,  m_Z^2} {\sf E_\nu} + {\sf N}_\nu  \right),\varrho \right] \nonumber \\
&+& \frac{x \, \widehat{C}[\varrho] }{m \, \overline{H}}\,, \label{eq:eomrho} \\
i \frac{d\bar\varrho}{dx}  &=& - \frac{x^2}{2 m^2\,y\, \overline{H}} \left[{\sf M}^2, {\bar\varrho} \right]
+ \frac{\sqrt{2} G_F \, m^2}{x^2 \, \overline{H}} \left[\left(+\frac{8\, y\, m^2}{3\, x^2 \,  m_W^2} {\sf E_\ell}+\frac{8\, y\, m^2 }{3 \, x^2 \, m_Z^2} {\sf E_\nu} + {\sf N}_\nu \right), {\bar\varrho} \right] \nonumber \\
&+& \frac{x \, \widehat{C}[{\bar\varrho}]}{m \, \overline{H}}\,, \label{eq:eombarrho}
 \,\,\,\,\,\,\,\,\,\,\,\,\,\,\,\,\,\,\,\,\,\,\,\,\,\,\,\,\,\,\,\,\,\,\,\,\,\,\,\,\,\,\,\,\,\,\,\,\,\,\,\,\,\,\,\,\,\,\,\,\,\,\,\,\,\,\,\,\,\,\,\,\,\,\,\,\,\,\,\,\,\,\,\,\,\,\,\,\,\,\,\,\,\,\,\,\,\,\,\,\,\,\,\,\,\,\,\,\,\,\,\,\,\,\,\,\,\,\,\,\,\,\,\,\,\,\,\,\,\,\,\,\,\,\,\,\,\,\,\,\,\,\,\,\,\,\,\,\,\,\,\,\,\,\,\,\,\,\,\,\,\,\,\,\,\,\,\,\,\,\,\,\,\,\,\,\ 
\label{eq:eomz}
\end{eqnarray}
which has to be solved along with the covariant conservation of the total stress-energy tensor
\begin{equation}
x \frac{d \varepsilon}{d x} = \varepsilon - 3 \P \, .
\end{equation}
In the previous expressions $\overline{H}$ denotes the properly normalized Hubble parameter
\begin{equation}
\overline{H} \equiv \frac{ x^2}{m} H = \frac{ x^2}{m} \sqrt{\frac{8\pi \, \epsilon(x,z(x))}{3 \, M_{Pl}^2}}=\left(\frac{m}{M_{Pl}}\right) \sqrt{\frac{8\pi \varepsilon(x,z(x))}{3}} \,,\label{hubbleeq}
\end{equation} 
and the total energy density and pressure enter through their ``comoving transformed'' values $ \varepsilon \equiv  \epsilon(x/m)^4 \simeq \varepsilon_{\gamma}+\varepsilon_{e}+\varepsilon_{\nu}$ and $\P \equiv P(x/m)^4$.  Compared with the treatment in~\cite{Mirizzi:2012we}, we now take into account the non-relativistic transition of the electrons and the entropy-transfer to the photons, responsible for increasing the temperature of photons with respect to the one of neutrinos, i.e. for making $z(x)> 1$. However, since we are only interested to situations where the change with respect to the instantaneous decoupling limit value $N_{\rm eff}=3$ is much larger than the $\sim 1.5\%$ effect due to incomplete neutrino decoupling during $e^{+}-e^{-}$ annihilation~\cite{Mangano:2005cc}, we neglect the  latter effect. We can thus track $z(x)$ once and for all by using the integrated entropy ratio formula (see e.g.~Eq. (15) in~\cite{Esposito:2000hi}, with $df_i/dx$ at the right-hand side put to zero).

The first term at the right-hand side of Eqs.~(\ref{eq:eomrho})-(\ref{eq:eombarrho}) accounts for vacuum oscillations, where in the flavour basis
 ${\sf M}^2 = {\mathcal U}^{\dagger}{\mathcal M}^2{\mathcal U}$. Here 
${\mathcal U}={\mathcal U}(\theta_{e\mu}, \theta_{es}, \theta_{\mu s})$
is the $3 \times 3$ active-sterile mixing matrix, parametrized as in~\cite{Mirizzi:2012we}.    We assume $\theta_{e\mu}$ equal to the active $1-3$ 
mixing angle $\theta_{13}$~\cite{Fogli:2012ua},   while we fix the active-sterile mixing angles to the best-fit values of the different 
anomalies~\cite{Giunti:2011gz}
\begin{eqnarray}
\sin^2  \theta_{e\mu} &=& 0.024  \,\ , \\
\sin^2 \theta_{es}  &=& 0.025 \,\ ,  \\
\sin^2 \theta_{\mu s}  &=& 0.023 \,\ .
\label{eq:stermix}
\end{eqnarray}
The mass-squared  matrix ${\mathcal M}^2 = \textrm{diag}(-\Delta m^2_{\rm atm}/2,
+\Delta m^2_{\rm atm}/2,\Delta m^2_{\rm st})$ is parametrized in terms of the atmospheric mass-squared difference
$\Delta m^2_{\rm atm}= 2.43 \times 10^{-3}$~eV$^2$~\cite{Fogli:2012ua} and of the active-sterile mass splitting 
$\Delta m^2_{\rm st}=0.89$~eV$^2$, fixed from the short-baseline fit in 3+1 model~\cite{Giunti:2011gz}. In the following
we assume the normal mass hierarchy $\Delta m^2_{\rm atm} >0$.
We checked that results similar to the ones we will present would have been obtained considering the (2+1) sub-sector associated with
the solar mass squared difference $\Delta m^2_{\rm sol}$ and with the $1-2$ mixing angle $\theta_{12}$.

The terms proportional to $G_F$ in  Eqs.~(\ref{eq:eomrho})-(\ref{eq:eombarrho}) encode the matter effects in the neutrino oscillations.
In particular, the term ${\sf E_\ell}$ is related to the energy density of electrons and positrons
(see Eq.~(20) in~\cite{Mirizzi:2012we}), 
while the $\nu-\nu$ interaction  terms are given by 
 \begin{eqnarray}
 {\sf N}_\nu &=&  \frac{1}{2 \pi^2} \int  dy \,  y^2 \,
 \{ {\sf G}_s (\varrho(x,y) -{\bar\varrho}(x,y)){\sf G}_s + {\sf G}_s \textrm{Tr} \left[(\varrho(x,y) -{\bar\varrho}(x,y)){\sf G}_s\right]  \}
 \,,\label{eq:neutrinodens}  \\
  {\sf E}_\nu& = &  \frac{1}{2 \pi^2} \int   dy \, y^3 \,
 {\sf G}_s (\varrho(x,y) +\bar\varrho(x,y)){\sf G}_s  \,.
 \label{eq:neutrinodens2}
 \end{eqnarray}
 Note that the matrix ${\sf N}_\nu$ is related to the {\it difference} of the  density matrices of neutrinos and antineutrinos, while ${\sf E}_\nu$ is related to their sum. The matrix ${\sf G}_s =\textrm{diag}(1,1,0)$  in flavour space contains the dimensionless coupling constants. We remark that considering more than one active species,  the ${\sf N}_\nu$ matrix also contains off-diagonal terms. 
In the presence of large neutrino asymmetries among the active species, as the ones we are considering in this work, the
matter effects are dominated by  the ${\sf N}_\nu$ contribution. This latter term makes  the EoMs non-linear and are the main numerical challenge in dealing with this physical system.
At large temperatures it dominates over the vacuum term, suppressing any flavour conversions. However, as the universe expands a \emph{resonance} condition  can be satisfied at lower temperatures, as extensively illustrated in~\cite{Mirizzi:2012we}.  

Finally, the last term in  Eqs.~(\ref{eq:eomrho})-(\ref{eq:eombarrho})
is the collisional one proportional to $G_F^2$, for which we adopt the same approximate expression $\widehat{C}[{\rho}]$  as in Eq. (28)
 of~\cite{Mirizzi:2012we}, but  now  keeping the $y-$dependence, see~\cite{Bell:1998ds}.
This guarantees that: i)   the correct collisional term is recovered when integrating the EoMs for $\varrho(y)$ over momenta,  ii)  the  overall lepton number conservation is  preserved. Note that this is not the case for alternative damping prescriptions often found in the literature, where the lepton number conservation is achieved imposing
an additional equation (see e.g.~\cite{Enqvist:1990ad}). 
We remark that possible minor inaccuracies in the $y-$dependence of the collisional terms are of little concern for our application, since 
in the cases where large spectral distortions are produced  in the electron (anti)neutrino distributions, these are caused by oscillatory terms rather than collisional ones. In other words, whenever collisional repopulation of depleted active neutrinos is relevant, one has very little departures from thermal spectra, while in the opposite limit, the major distortion is not due to the exact $y-$dependence of the collisional terms. On the other hand, implementing an operator enforcing lepton number conservation (numerically to a very high degree
of precision) is crucial
to avoid spurious numerical effects.

In order to fix the initial conditions for the flavour evolution, we notice that 
active neutrinos are produced in the very early universe with their energy
spectrum  kept in chemical and kinetic equilibrium by weak interactions
until temperatures $T \simeq \,$few MeV,
 when the corresponding collision
rates fall below the cosmological expansion rate. 
In the presence of primordial neutrino asymmetries, the original 
active neutrino spectra are given by 
Fermi-Dirac distributions
 parametrized in terms of a temperature $T_\nu$ and 
 chemical potentials $\mu_{\alpha}$
for $\alpha=e,\mu,\tau$. Each 
 neutrino asymmetry in Eq.~(\ref{eq:leptonasy})  can be expressed
in terms of the corresponding degeneracy parameter $\xi_{\alpha}= \mu_{\alpha}/T_\nu $ as
\begin{equation}
L_{\alpha} = \frac{1}{12 \zeta(3)}\left(\frac{T_{\nu}}{T_\gamma}\right)^3
(\pi^2 \xi_{\alpha} + \xi_{\alpha}^3) \simeq 0.68 \,\ \xi_{\alpha} \left(\frac{T_{\nu}}{T_\gamma}\right)^3 \,\ ,
\end{equation}
with $\zeta(3)\simeq1.202$, where the right hand side numerical expression corresponds to 
the leading order in small $ \xi_{\alpha}$. 
In the rest of the paper,  we shall indicate the neutrino asymmetries in terms of the $\xi_{\alpha}$ parameters rather than $L_{\alpha}$, in order to conform with the more frequently used notation in phenomenological
papers.

The initial conditions for the  density matrix $\varrho$ are then given by
\begin{eqnarray}
\varrho_{\rm in} &=& \textrm{diag}\left(f_{\rm eq} (y,\xi_e), f_{\rm eq} (y,\xi_\mu), 0\right) \,\ , \nonumber \\
{\bar\varrho} _{\rm in} &=& \textrm{diag}\left(f_{\rm eq} (y,-\xi_e), f_{\rm eq} (y,-\xi_\mu), 0\right)
\label{43} \,\ ,
\end{eqnarray}
with $f_{\rm eq} (y,\xi) = 1/[\textrm{exp}(y-\xi)+1]$.

 We remark that since in our study we consider initial distributions for active neutrinos close to their equilibrium ones~\footnote{This is conservative since some population of sterile
neutrinos cannot be excluded in extensions of the Standard Model, following for example from decays of heavier neutrino singlet states.}, the
oscillations among the three active species have a qualitatively important role but quantitatively their details are of minor relevance for the evolution of the sterile neutrinos.
Therefore, the (2+1) scenario we consider is a good proxy for the complete (3+1) scenario.
We also mention that in a recent work~\cite{Hannestad:2012ky}, it has been performed a multi-momentum study 
of the kinetic equations for a system consisting of only one  active and one sterile neutrino species [i.e. (1+1) scenario].
 While instructive in several respects, this is quite a simplified
scenario since it does not allow to incorporate effects like the existence of more than one active-sterile mixing angle and different choices of the neutrino asymmetries
for the different flavours.

\subsection{Results}\label{multi}

In this Section we present our results for  the sterile neutrino abundance in the (2+1) scenario described in Sec.~\ref{formalism}.
 We numerically
solve the EoMs Eqs.~(\ref{eq:eomrho})-(\ref{eq:eombarrho}) with an integration routine for stiff ordinary differential equations
taken from the NAG libraries~\cite{nag}  and based on an adaptive method. The range for $x$ is chosen to be $x \in [2\times 10^{-2},0.5]$. 
As a compromise between energy resolution of the spectral distortions and computational cost, we took $N_y = 21$ momentum modes
in the range $y \in [0,10]$.
The grid points are not chosen to be equally spaced, but are instead fixed by
imposing weighted Gaussian quadrature of the  integrals in the right-hand-side
of Eqs.~(\ref{eq:neutrinodens})--(\ref{eq:neutrinodens2}). By increasing the momentum grid points to  $N_y = 30$, we checked in some test runs that this is enough to keep the error below the per-cent level on the effective number of neutrinos $\neff$.

\begin{figure}
\begin{center}
 \includegraphics[angle=0,width=0.7\textwidth]{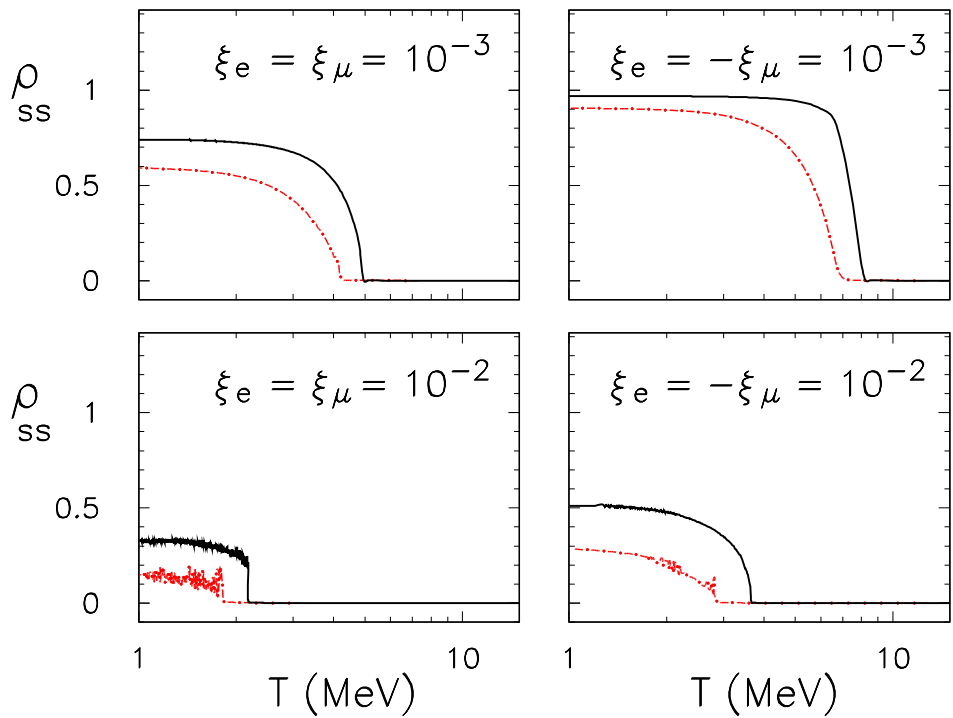} 
    \end{center}
\caption{\label{fig2n} Evolution of the total value of the sterile neutrino density matrix element $\rho_{ss}$ 
as function of the temperature $T$ for the multi-momentum case (continuos curves) and the average momentum  (dot-dashed curves) case with 
a thermal momentum $\langle y \rangle=3.15$. Left panels correspond to $\xi_e=\xi_\mu$, while in the
lowers $\xi_e=-\xi_\mu$. Upper panels refer to $\xi_e=10^{-3}$ and right to  $\xi_e=10^{-2}$.} 
\end{figure}

We remark that due to the momentum dependence of the resonance conditions,
in the multi-momentum treatment of the EoMs there can be significant deviations with respect to
 the evolution predicted by the average momentum scheme. A direct comparison between the single-momentum and the multi-momentum results is reported
 in Fig.~\ref{fig2n}, which shows: 
\emph{(a)} 
the 
momentum-integrated sterile neutrino density matrix element (solid curves) normalized to the integral of a Fermi-Dirac distribution with zero chemical potential,
\begin{equation}
\rho_{ss}(x) = \frac{\int dy \,\ y^2 \varrho_{ss}(x, y)}{\int dy \,\ y^2 f_{\rm eq} (y,0)} \,\ ,
\end{equation}

\emph{(b)} the  sterile neutrino density matrix element $\rho_{ss}$
in the average momentum scheme (dot-dashed curves), normalized correspondingly.
 In the left panels we take equal initial neutrino asymmetries, 
$\xi_e=\xi_\mu$, while in the right panels we refer to opposite ones,
$\xi_e=-\xi_\mu$. In the upper panels we consider
$\xi_e=10^{-3}$, while in the lower panels we take 
$\xi_e=10^{-2}$.
In all the considered cases the values of $\rho_{ss}$
for the single momentum underestimates the sterile neutrino abundance with respect to the 
multi-momentum case. The enhancement obtained with the multi-momentum treatment is significant, roughly
$\sim 20 \%$ of a degree of freedom. 
Moreover, we note that the sterile production in the multi-momentum case occurs at higher temperatures
with respect to the average momentum case. This is due to the fact that in the multi-momentum evolution the 
sterile neutrino population can start building up earlier via lower momenta modes, that resonate earlier than the average momentum.
This anticipates the dynamical evolution of $\xi$  and is the main responsible for the difference observed. Also, we confirm the observation
of~\cite{DiBari:2001jk} that the resonance is more adiabatic  at higher temperature.
Hence, the average momentum treatment of the EoMs is generically expected to {\it underestimate} the sterile neutrino abundance.

\section{Impact on observables: $N_{\rm eff}$ and light nuclei abundances}\label{BBN}
 Phenomenological
quantities affected by active-sterile neutrino flavour conversions notably  depend on the overall non electromagnetic radiation content, parametrized via $N_{\rm eff}$, and from
the distortions of the electron (anti)neutrino spectra, the latter being a basic input for BBN weak rates~\footnote{Actually, treating properly effects sensitive to the neutrino masses require knowing the flavour composition of the neutrino ensemble as well. Similarly, precision computations of CMB anisotropies are in principle sensitive to the neutrino
phase-space distributions, see e.g. ~\cite{Lesgourgues:2011rh}.}.

In the upper panels of Figs.~2 and~3 we show
 the  $y$-dependent  $\nu_e$ 
energy spectrum $y^2 \varrho_{ee}(y)$ (dashed curve) at $T=1$~MeV, compared with the initial one 
$y^2 f_{\rm eq}(y,\xi_e)$ (solid line). 
 In particular Fig.~2 refers to $\xi_e = 10^{-3}$ and Fig.~3 to
 $\xi_e = 10^{-2}$. In each Figure, in the left panels $\xi_e = \xi_\mu$, while in the right ones $\xi_e = -\xi_\mu$.
 In order to characterize the distortion in the $\nu_e$ spectra with respect to the initial one, in the lower panel we plot
the ratio 
\begin{equation}
R = \frac{\varrho_{ee}(y)}{f_{\rm eq}(y,\xi_e)} \,\ .
\end{equation}
In the case of $\xi_e = 10^{-3}$, $R \gtrsim 0.95$ for equal asymmetries and $R \gtrsim 0.98$ in the 
case of opposite asymmetries.
Conversely the spectral distortions are more evident for  $\xi_e = 10^{-2}$. Namely, for equal asymmetries one finds
$R \gtrsim 0.82$, while for opposite asymmetries  $R \gtrsim 0.9$.
Indeed, spectral distortions in the active neutrinos are more evident when resonant active-sterile conversions
occur near the active neutrino decoupling temperature, as pointed out at first in~\cite{Abazajian:2004aj}.  

We remind that, as already commented in~\cite{Mirizzi:2012we}, the dynamical
evolution of the asymmetries is such that {\it both} neutrinos and antineutrinos get populated resonantly, roughly in equal values. Hence, 
we only show here the results referring to neutrinos. Of course, in the numerical computation the small differences between the two sectors have been properly
accounted for.

Concerning the effective number of neutrino species, we remark that at the level of approximation we are adopting, $N_{\rm eff}=3+\Delta N_{\rm eff}$ enters the dynamics only via its contribution to the Hubble parameter, see Eq.~(\ref{hubbleeq}), 
where it rescales the standard neutrino energy density contribution $\varepsilon_\nu$ as 
\begin{equation}
\varepsilon_\nu(x,N_{\rm eff})\to \varepsilon_\nu(x,3)\left(1+\frac{\Delta N_{\rm eff}}{3}\right)\,.
\end{equation} 
Technically, we compute $\Delta N_{\rm eff}$ in the above equation via the following (numerical) integral
\begin{equation}
\Delta N_{\rm eff} = \frac{60}{7 \pi^4} \int dy \,\ y^3 
\textrm{Tr}[\varrho(x,y) +\bar\varrho(x,y)] \,\ - 2 \,\ ,\label{deltaneff}
\end{equation}
the factor ``$-2$''  being due to the fact that we are considering only two active neutrino species.

The quantity defined in Eq.~(\ref{deltaneff}) is shown in Figure~\ref{fig5n} for two representative values of asymmetries, 
($\xi_e = 10^{-3}$ for solid curves and $\xi_e = 10^{-2}$ for dashed curves), taken equal (opposite) for the $e$ and $\mu$ sector 
in the left (right) panel.  
For $\xi_e= 10^{-3}$, we see that the resonant production of sterile neutrinos starts around $T \simeq 5$~MeV for equal
asymmetries and $T \simeq 8$~MeV for opposite ones. Since active-sterile neutrino conversions mostly occur when the collisional regime is still operative, the active neutrino species are almost fully repopulated,
reflecting in a $\Delta N_{\rm eff} \simeq \rho_{ss}$ in both cases (see Fig.~1).  This is still true to some extent for the case $\xi_e=-\xi_\mu =10^{-2}$, but not when  $\xi_e=\xi_\mu =10^{-2}$. Namely,
in this latter situation, the resonant population starts at temperatures as low as $T\sim 2\,$MeV, comparable to neutrino decoupling temperature. This means that active neutrino repopulation
 is only partial.  Also, an appreciable difference (in this case of $\sim 0.1$) is established between $\rho_{ss}$ and $\neff$. These effects were already noted in~\cite{Mirizzi:2012we}, where they were even more prominent due to
the ``less effective'' sterile production associated to the single momentum approximation. 

\begin{figure}[!tb]
\begin{center}
 \includegraphics[angle=0,width=0.6\textwidth]{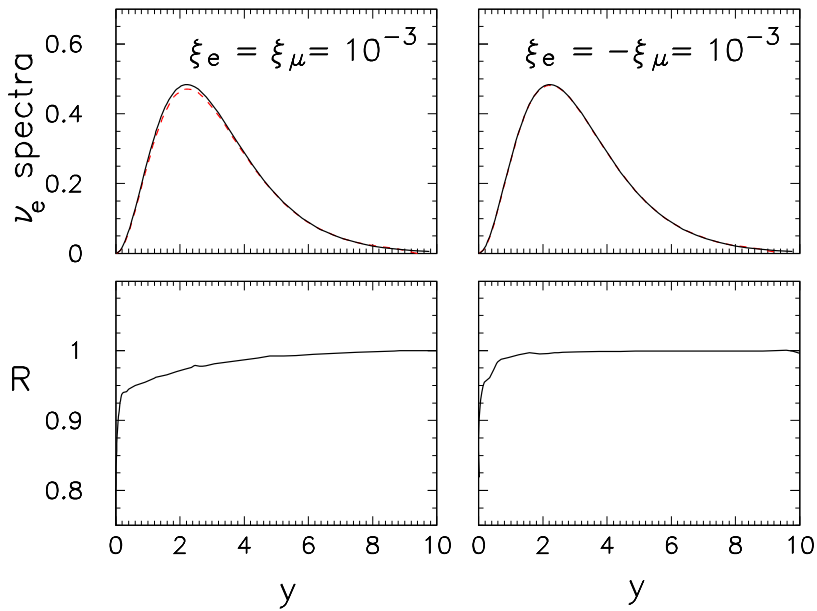} 
    \end{center}\label{fig3n}
\caption{Cases with $\xi_{e}= 10^{-3} $. \emph{Upper panels}: Final  $\nu_{e}$ energy spectra  at $T= 1$~ MeV   (dashed curve) and initial ones  (continuous curve). \emph{Lower panel}: Ratio $R$ between final and initial $\nu_{e}$ energy spectra. Left panels refer to $\xi_{e}= \xi_{\mu}$ while right panels are for  $\xi_{e}= -\xi_{\mu}$.} 
\end{figure}

\begin{figure}[!t]
\begin{center}
 \includegraphics[angle=0,width=0.6\textwidth]{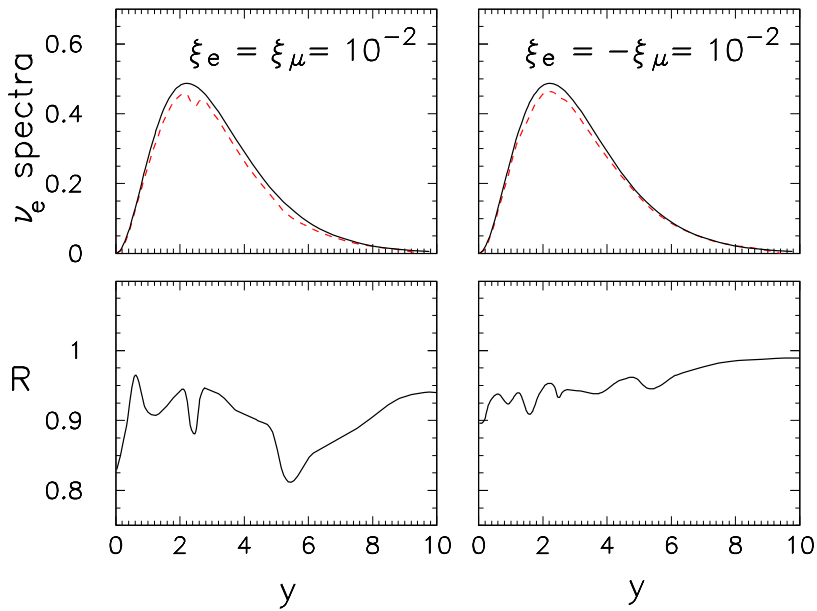} 
    \end{center}\label{fig4n}
\caption{Cases with $\xi_{e}= 10^{-2} $. \emph{Upper panels}: Final  $\nu_{e}$ energy spectra  at $T= 1$~ MeV   (dashed curve) and initial ones  (continuous curve). \emph{Lower panel}: Ratio $R$ between final and initial $\nu_{e}$ energy spectra. Left panels refer to $\xi_{e}= \xi_{\mu}$ while right panels are for  $\xi_{e}= -\xi_{\mu}$.} 
\end{figure}

The numerical values of the $\Delta N_{\rm eff}$'s found in a few representative runs are reported in Table~\ref{tab1} along with the values of the yields of  $^4$He mass fraction
$Y_p$
 and deuterium $^2$H, as obtained from a modified version of the numerical code \texttt{PArthENoPE}~\cite{Pisanti:2007hk} for a baryon fraction $\omega_b=0.02249$ and the neutron lifetime $\tau_n=880.1\,$s, following Particle Data Group 2012 recommendations~\cite{Beringer:1900zz}. For comparison we also show  cases with neutrino asymmetries but no sterile neutrinos as well as  the standard BBN case.

Technically, note that the rates $\Gamma_{n\to p}[f_{\nu_e},f_{\bar\nu_e}]$ and $\Gamma_{p\to n}[f_{\nu_e},f_{\bar\nu_e}]$ are functionals of the 
 distributions $f_{\nu_e},f_{\bar\nu_e}$. If we denote with $\Gamma^0$ the rates computed in the Born approximation  for Fermi-Dirac spectra,
  and with  $\Gamma$ the actual rates for the cases at hand,  we have computed the effect of sterile neutrinos by rescaling the rates implemented in the code~\texttt{PArthENoPE}~\cite{Pisanti:2007hk} (see also~\cite{Serpico:2004gx}) by $\Gamma/\Gamma^0$, which has been numerically evaluated and then interpolated.
This amounts to a first-order correction in a perturbative approach. However, since the rate corrections is at most ${\cal O}(3\%)$ for the largest asymmetries,
the error due to this approximation is safely below ${0.3\%}$ (see e.g. \cite{Esposito:1998rc} for the analysis of corrections to Born weak rates). Thus, this is comparable or lower than neglecting the modification to the reheating in the standard scenario. 
In determining the shape of the distribution function, especially for the cases with $|\xi|=10^{-2}$ the main source of error comes in fact from the discretization of the neutrino distribution (and the corresponding interpolation). Still, test runs with a $N_y = 30$ points grid in momentum space suggest that this does not spoil the reliability of the size of the effects we found.

With reference to Table~\ref{tab1}, a few comments are in order. First, note that for sufficiently small values of $\xi_{e,\mu}$, all effect of sterile states on BBN is due to the increased $N_{\rm eff}$, which is in any case larger than what found in the single momentum approximation for the same value of $|\xi|$. This holds true for both $Y_p$ and \h2. On the other hand, for high values, say $|\xi_{e,\mu}| \sim 10^{-2}$, some fraction of the sterile neutrino population builds-up relatively late, namely {\it after} the freeze-out of the active neutrinos. These cases  are associated to a $\Delta N_{\rm eff}$ significantly smaller than 1. While this quantity is still the main cause of the change in \h2, a significant fraction of the effect on $Y_p$ is due to the changes of the weak rates regulating the $n\leftrightarrow p$ chemical equilibrium due to distorted $\nu_e$ and $\bar{\nu}_e$ distributions. In the case $\xi_e=-\xi_\mu=10^{-2}$, this effect is $\sim 75$\% than the one related to the speed-up of the expansion due to $\Delta N_{\rm eff}$, while for  $\xi_e=\xi_\mu=10^{-2}$ it  becomes three times larger than the other~\footnote{The evolution for this latter case is extremely slow and there is still a small evolution in the parameters taking place at $T\lesssim 1\,$MeV. Hence the results presented here, which assume that the asymptotic results are equal to the ones at the smallest temperatures followed  are slightly {\it conservative}. The actual effect should be a bit larger.}.
 Phenomenologically, these  scenarios  with large asymmetries in the presence of sterile neutrinos yield comparatively
 lower values of $N_{\rm eff}$ as probed by CMB (say,  $N_{\rm eff} \simeq 3.2$), while altering $Y_p$ by an amount loosely equivalent
to a larger $N_{\rm eff}$. We note that in these cases the presence of eV sterile neutrinos pushes the change of $Y_p$ in the direction
of an \emph{increase} with respect to the standard BBN value. This behavior confirms the prediction based on analytical estimations 
presented in~\cite{Smith:2006uw}.  
 The comparison with the last three rows in Table~\ref{tab1} also shows that this trend is {\it opposite} of what obtained
in presence of a positive-sign electron neutrino asymmetry (in absence of sterile neutrinos).
Actually, in some previous phenomenological analyses, not treating sterile neutrinos dynamically  (like in~\cite{Hamann:2011ge}) it has been envisaged the possibility ``to mask''  the presence of extra eV-scale sterile neutrino degrees of freedom to BBN by introducing large chemical potentials,
treating the system as a ``degenerate BBN'' plus extra radiation. 
This prescription may
lead to even {\it qualitatively wrong} conclusions, such as the positive correlation in allowed regions between the increase of $\xi$ and $N_{\rm eff}$,
visible e.g. in Fig.~6 of Ref.~\cite{Hamann:2011ge}.

Quantitatively, we observe that the modifications of BBN yields induced by sterile neutrinos are sizable, as reported in Table I.
For comparison, the statistical error on the astrophysical $Y_p$ determination
can be as small as 0.001 (albeit the systematic error
is at the moment several times larger)~\cite{Izotov:2010ca}, the
determination of ${}^2{\rm H}/{\rm H}$ in the highest quality Quasar system
is $(2.535 \pm 0.05) \times 10^{-5}$~\cite{Pettini:2012ph}, and the $1 \sigma$ errors  on $N_{\rm eff}$ and $Y_p$ reported by the combined analysis of the Planck team
amount to $\Delta N_{\rm eff}\simeq 0.27$ and $\Delta Y_p \simeq 0.021$~\cite{Ade:2013lta}, in substantial agreements with earlier forecasts~\cite{Perotto:2006rj}.

Finally, we checked that  changing squared mass differences and mixing angles within current uncertainties~\cite{Giuntipriv},  one can easily obtain ${\cal O}(10\%)$ differences in $\neff$.  Also, the results presented here are rather on the conservative side, as far as the particle mass parameter is chosen. For example, considering only disappearance experiments,
larger values of sterile mass splitting than the 0.89 eV$^2$ used here are preferred, see e.g.~\cite{Giunti:2012bc}. In this case, the results we obtained are modified in the sense of an easier thermalization of the sterile state.


\begin{figure}
\begin{center}
 \includegraphics[angle=0,width=0.8\textwidth]{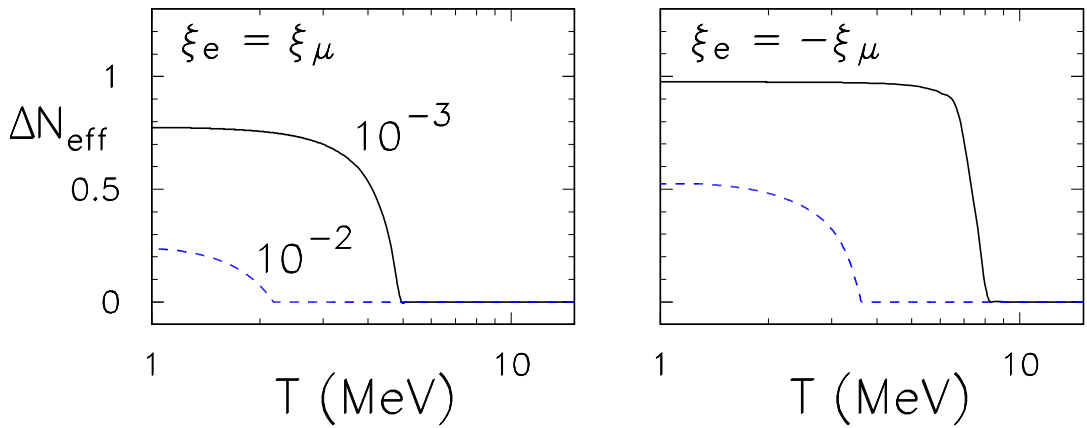} 
    \end{center}
\caption{\label{fig5n} Evolution of  $\Delta N_{\rm eff}$  vs. temperature $T$
for equal asymmetries $\xi_e=\xi_\mu$ (left panel) and opposite asymmetries $\xi_e=-\xi_\mu$
 (right panel). Solid curves refer to $\xi_e=10^{-3}$ while dashed curves are for 
 $\xi_e=10^{-2}$.
} 
\end{figure}

%
{\begin{table}[t]
\begin{tabular}{||l||r|r|r|r||}\hline
Case & $\Delta \neff$& $\Delta \neff^{\langle y\rangle}$ & $Y_p$ & \h2/H ($\times 10^5$)   \\
\hline\hline 
$|\xi|\ll 10^{-3}$ & 1.0 & 1.0 & 0.259
& 2.90 \\
\hline
$\xi_e=-\xi_\mu=10^{-3}$ & 0.98 & 0.89 & 0.257
& 2.87  \\
\hline
$\xi_e=\xi_\mu=10^{-3}$ & 0.77 & 0.51 & 0.256
& 2.81  \\
\hline
$\xi_e=-\xi_\mu=10^{-2}$& 0.52 & 0.44 & 0.255
& 2.74  \\
\hline
$\xi_e=\xi_\mu=10^{-2}$ & 0.22 & 0.04 & 0.251 
& 2.64  \\
\hline\hline
$\xi_e=|\xi_\mu|=10^{-3}$, no $\nu_s$ & $\sim$ 0 & -- & 0.246 
& 2.56  \\
\hline\hline
$\xi_e=|\xi_\mu|=10^{-2}$, no $\nu_s$ & $\sim$ 0 & -- & 0.244 
& 2.55  \\
\hline\hline
standard BBN & 0 & 0 & 0.247
& 2.56  \\
\hline
\end{tabular}
\caption{The values of $\Delta \neff$ and the calculated abundances of $^4$He mass fraction
$Y_p$
 and deuterium $^2$H in the different cases considered in this paper.  For comparison, the third column refers to the increase in the effective degrees of freedom obtained in the average momentum approximation, $\Delta \neff^{\langle y\rangle}$.}\label{tab1}
\end{table}
%

\section{Conclusions}\label{concl}
In this article, we have presented for the first time the results of a multi-flavour, multi-momentum computation of cosmological neutrino spectra in presence
of a sterile state with parameters consistent with those invoked in the  interpretations of short-baseline neutrino oscillation
experiments~\cite{Giunti:2011gz}. We have considered the case of relatively
large neutrino asymmetries ($|L_{\nu}|\gtrsim 10^{-3}-10^{-2}$), which suppress the sterile  production in the early universe giving in this way a better agreement with several cosmological observations. 
Our results show an enhancement  in the sterile neutrino abundance of up to 0.2 effective degrees of freedom with respect to what observed in the average momentum 
study. This shifts the asymmetries needed 
for a significant suppression of $\neff$ to relatively larger values, of the order of $|L_{\nu}|\gtrsim 10^{-2}$.  Moreover, starting with opposite asymmetries in different flavours with  vanishing  net neutrino asymmetries provides a less effective inhibition. 
On the other hand, 
for large asymmetries the significant production of the sterile neutrinos after the $\nu_e$ and $\bar{\nu}_e$ decoupling causes non-negligible distortions 
(from few $\%$ up to $\sim 20\%$ for some cases considered) in the $\nu_e$ and $\bar{\nu}_e$ energy distributions. 
As a result---leaving apart the  problem of how to  generate such large initial asymmetries---the modifications of BBN yields are sizable, as reported in Table~\ref{tab1}. 
Also worth noting, while for $^2$H/H these modifications are essentially due to the larger $\neff$, for the $^4$He mass fraction $Y_p$ a significant effect follows from (anti) neutrino distribution distortions produced in presence of both asymmetries and a sterile state. 

Although it is not the purpose of the present paper to provide cosmological constraints on sterile neutrinos, we can outline  the two following possible scenarios: \\

i) 
thermalization of the sterile state in the presence of small or vanishing neutrino asymmetries. In the ``interesting'' parameter space this  is often close to total, and already with  pre-Planck data tensions were manifest between cosmological disfavored regions and laboratory claims (see e.g. Fig. 5 in~\cite{Melchiorri:2008gq} or the left panel of Fig. 3 in~\cite{Archidiacono:2012ri}). 
This is amenable to a systematic analysis of the sterile neutrino mass and mixing parameter space, and a first analysis following the Planck data release is strengthening the above-mentioned tensions~\cite{Mirizzi:2013kva}. \\

ii) No complete thermalization of sterile neutrinos. Extra ingredients (which imply major cosmological changes) must be introduced.
  The most popular proposal in the literature has been  to invoke neutrino asymmetries, whose effects on cosmological
observables has been treated with simplified recipes. Here we put this possibility under close scrutiny, finding notable differences with naive expectations found in the literature.
This scenario could lead to a possible inconsistency in the value of $N_{\rm eff}$ extracted from CMB and BBN.
Indeed, sterile neutrinos and large asymmetries would produce 
a
relatively low value of $N_{\rm eff}$ as probed by CMB, and an  increase of $Y_p$.
This latter
would be mimicked by   a {\it larger} $\neff$.  It is also worth stressing that the standard BBN prediction $Y_p= Y_p(\omega_b,\neff)$ {\it is not valid anymore}. 
Given our current CPU power,  we have limited our analysis  of this scenario to few representative cases. Contrarily to what proposed in the scenario 
i) 
a large scan over parameter space (which would be needed for a quantitative statistical analysis combining CMB, LSS and BBN data) appears a hard  task. 
Indeed, any precise quantitative result for the cosmological observables would depend on an interplay of active-active and active-sterile neutrino oscillations (multi-flavour effects), whose matrix structure is not  fully determined by current data on short-baseline neutrino anomalies (most notably in the $\tau-s$ sector).  Hence ``precision'' computations would be illusory and premature, given the dependence from unknown or poorly constrained parameters. 

Future directions of these studies would depend on the fate of the short-baseline neutrino anomalies and on the current generation of cosmological measurements. Definitely, these results are already orienting---and will orient even more in the near future---further research on the non-trivial role played by light sterile neutrinos in cosmology.

\section*{Acknowledgements} 

We thank Carlo Giunti for providing us preliminary results from new fits of the short-baseline 
neutrino data. We acknokledge Irene Tamborra for careful reading the draft and for useful comments on it.
The work of A.M. and N.S.  was supported by the German Science Foundation (DFG)
within the Collaborative Research Center 676 ``Particles, Strings and the
Early Universe.'' G. M, G. M, and O.P  acknowledge support by
the {\it Istituto Nazionale di Fisica Nucleare} I.S. FA51 and the PRIN
2010 ``Fisica Astroparticellare: Neutrini ed Universo Primordiale'' of the
Italian {\it Ministero dell'Istruzione, Universit\`a e Ricerca}. 
G. Miele and O.P. acknowledge
partial support by Progetto FARO 2010 of the University of Naples
Federico II. At LAPTh, this activity was developed coherently with the research axes supported by the excellence laboratory (Labex) grant ENIGMASS. 


\end{document}